\documentclass[aps,reprint, prl, superscriptaddress, showpacs,nobibnotes]{revtex4-1}
\usepackage{graphicx}

\begin{document}

\title{Resolving the nature of electronic excitations in resonant inelastic x-ray scattering}%

\author{M. Kang}%
\thanks{These authors contributed equally}
\affiliation{Department of Physics, Massachusetts Institute of Technology, Cambridge, Massachusetts 02139, USA}
\author{J. Pelliciari}%
\thanks{These authors contributed equally}
\affiliation{Department of Physics, Massachusetts Institute of Technology, Cambridge, Massachusetts 02139, USA}
\author{Y. Krockenberger}%
\affiliation{NTT Basic Research Laboratories, NTT corporation, 3-1 Morinosato-Wakamiya, Atsugi, Kanagawa 243-0198, Japan.}
\author{J. Li}%
\affiliation{Department of Physics, Massachusetts Institute of Technology, Cambridge, Massachusetts 02139, USA}
\author{D. E. McNally}%
\affiliation{Research Department Synchrotron Radiation and Nanotechnology, Paul Scherrer Institut, 5232 Villigen PSI, Switzerland.}
\author{E. Paris}%
\affiliation{Research Department Synchrotron Radiation and Nanotechnology, Paul Scherrer Institut, 5232 Villigen PSI, Switzerland.}
\author{R. Liang}%
\affiliation{Department of Physics and Astronomy, University of British Columbia V6T 1Z1, Canada.}
\affiliation{Quantum Matter Institute, University of British Columbia V6T 1Z4, Canada.}
\author{W. N. Hardy}%
\affiliation{Department of Physics and Astronomy, University of British Columbia V6T 1Z1, Canada.}
\affiliation{Quantum Matter Institute, University of British Columbia V6T 1Z4, Canada.}
\author{D. A. Bonn}%
\affiliation{Department of Physics and Astronomy, University of British Columbia V6T 1Z1, Canada.}
\affiliation{Quantum Matter Institute, University of British Columbia V6T 1Z4, Canada.}
\author{H. Yamamoto}%
\affiliation{NTT Basic Research Laboratories, NTT corporation, 3-1 Morinosato-Wakamiya, Atsugi, Kanagawa 243-0198, Japan.}
\author{T. Schmitt}%
\affiliation{Research Department Synchrotron Radiation and Nanotechnology, Paul Scherrer Institut, 5232 Villigen PSI, Switzerland.}
\author{R. Comin}%
\email{rcomin@mit.edu}
\affiliation{Department of Physics, Massachusetts Institute of Technology, Cambridge, Massachusetts 02139, USA}

\begin{abstract}
The study of elementary bosonic excitations is essential toward a complete description of quantum electronic solids. In this context, resonant inelastic X-ray scattering (RIXS) has recently risen to becoming a versatile probe of electronic excitations in strongly correlated electron systems. The nature of the radiation-matter interaction endows RIXS with the ability to resolve the charge, spin and orbital nature of individual excitations. However, this capability has been only marginally explored to date. Here, we demonstrate a systematic method for the extraction of the character of excitations as imprinted in the azimuthal dependence of the RIXS signal. Using this novel approach, we resolve the charge, spin, and orbital nature of elastic scattering, (para-)magnon/bimagnon modes, and higher energy \textit{dd} excitations in magnetically-ordered and superconducting copper-oxide perovskites (Nd$_{2}$CuO$_{4}$ and YBa$_{2}$Cu$_{3}$O$_{6.75}$). Our method derives from a direct application of scattering theory, enabling us to deconstruct the complex scattering tensor as a function of energy loss. In particular, we use the characteristic tensorial nature of each excitation to precisely and reliably disentangle the charge and spin contributions to the low energy RIXS spectrum. This procedure enables to separately track the evolution of spin and charge spectral distributions in cuprates with doping. Our results demonstrate a new capability that can be integrated into the RIXS toolset, and that promises to be widely applicable to materials with intertwined spin, orbital, and charge excitations.
\end{abstract}

\date{\today}%
\maketitle
%\tableofcontents

%%%%%%%%%%%%%%%%%%%%%%%%%Sec1

\section{I. Introduction}
The emergence of collective excitations associated with different, often coupled degrees of freedom, is a common trait in strongly interacting systems. The detailed nature of the fundamental interactions is reflected not only in the momentum-energy spectrum but also in the character of these emerging excitations. In recent years, RIXS has earned a leading role in the study of electronic excitations in quantum materials, thanks to improved energy resolutions enabling access to low energy excitations \cite{ament_resonant_2011,hill_observation_2008,tacon_intense_2011,dean_persistence_2013,lee_asymmetry_2014,ishii_high-energy_2014,dean_spin_2012,schlappa_spinorbital_2012,ament_theoretical_2009,haverkort_theory_2010,ghiringhelli_observation_2009,ament_theory_2011,schlappa_collective_2009,jia_persistent_2014,zhou_persistent_2013,pelliciari_intralayer_2016}. In addition to its distinctive features including elemental selectivity, bulk sensitivity, and compatibility with small samples or thin films, RIXS covers an extended kinematic range in energy and momentum, complementing prominent scattering techniques such as neutron scattering, electron energy loss spectroscopy, and Raman/Brillouin scattering \cite{ament_resonant_2011}. At the same time, RIXS is sensitive to a broad array of excitations arising from the spin, charge, orbital, and lattice degrees of freedom. However, the assignment of individual excitations is intricate and has sometimes been elusive, leaving the full potential of RIXS untapped \cite{guarise_anisotropic_2014,benjamin_single-band_2014,kanasz-nagy_resonant_2016,minola_collective_2015,kung_doping_2015,jia_using_2016,huang_raman_2016}.

The complexity of the interpretation of RIXS spectra has represented a constant challenge for experimental and theoretical RIXS studies on cuprate \cite{guarise_anisotropic_2014,benjamin_single-band_2014,kanasz-nagy_resonant_2016,minola_collective_2015,kung_doping_2015,jia_using_2016}. In cuprates, the evolution of the low energy excitations from the antiferromagnetic parent insulator to the carrier-doped superconductor is a key piece in the grand puzzle of high-temperature superconductivity \cite{dean_insights_2015}. RIXS has detected persisting spin excitations across the phase diagram of the cuprates, producing experimental evidences to be accounted for in magnetic pairing theories \cite{tacon_intense_2011,dean_persistence_2013}. However, the radiation-matter interaction is extremely complicated and allows the observation of several degrees of freedom whose nature is hard to disentangle experimentally leading to possible different interpretation of the experimental data \cite{guarise_anisotropic_2014}.
%At the early stage of research, where the low energy excitations in RIXS spectra have been considered in total as a magnetic excitation, RIXS detects the persisting magnon-like excitations throughout the phase diagram, producing experimental evidences to be accounted for in magnetic pairing theories \cite{tacon_intense_2011,dean_persistence_2013}. However, follow-up experiment and RPA spin susceptibility calculation detect the partial descrepancy of spin-wave picture along the nodal direction in metallic overdoped regime, suggesting the excitations observed in RIXS spectra might have more complex nature \cite{guarise_anisotropic_2014}.
Subsequently, the various interesting theories have been proposed: some calculations argue that RIXS actually probes the spin dynamical structure factor in doped cuprate (i.e. paramagnon excitation) \cite{kung_doping_2015,jia_using_2016}, while others suggest the incoherent particle-hole excitations arising from the band structure effect as an origin of the experimental RIXS spectra in metallic cuprates \cite{benjamin_single-band_2014,kanasz-nagy_resonant_2016}. The full validation of these scenarios rests on the experimental capability to separately track down the doping evolution of the spin and charge susceptibility.

In other materials, it is the orbital degrees of freedom that plays an essential role in determining the electronic ground state and low-lying excitations. This class of strongly-correlated systems includes orbital-ordered nickelates and manganites, Fe-based superconductors, cobaltates, and spin-orbit coupled 5$d$ oxides \cite{chen_modifying_2013,bisogni_the_2016,wilkins_direct_2003, zhou_persistent_2013,pelliciari_intralayer_2016,pelliciari_presence_2016,pelliciari_local_2017,satoh_excitation_2017,kim_magnetic_2012,calder_spin-orbit-driven_2016,lu_doping_2017}. In these materials, the orbital degrees of freedom are dynamically active (and often coupled to spin and charge), and contribute to the spectrum of low-energy excitations, thus complicating the interpretation of RIXS spectra. Most importantly, the polarization analysis, besides requiring additional experimental components, provides only limited information in these cases, since the scattering matrix for orbital excitations is typically more asymmetric (see Appendix A) than the charge and spin channels. These considerations underscore the importance to develop a systematic method to resolve the nature and character of individual excitations encoded in the RIXS spectra.

RIXS is a 2$^{nd}$ order interaction process governed by a polarization-dependent cross section which can be derived from the Kramers-Heisenberg formula \cite{ament_resonant_2011,haverkort_theory_2010}. Most importantly, in the RIXS process, the character of each excitation is uniquely imprinted onto a distinctive form of the scattering tensor, which is ultimately determined by the matrix elements of the interaction (electric dipole) operator. The scattering tensor can be partly resolved by measuring the RIXS signal as a function of the polarization of incident ($\sigma_{in}$/$\pi_{in}$ : perpendicular/parallel to the scattering plane) and scattered ($\sigma_{out}$/$\pi_{out}$) photon beams, for a total of four polarization channels ($\sigma_{in}$-$\sigma_{out}$, $\sigma_{in}$-$\pi_{out}$, $\pi_{in}$-$\sigma_{out}$, $\pi_{in}$-$\pi_{out}$). Full polarization analysis is therefore often insightful \cite{braicovich_momentum_2010,ishii_polarization-analyzed_2011,minola_collective_2015} but ultimately insufficient to resolve the full (3$\times$3) scattering tensor, especially in systems with complex orbital physics where all components are nonzero and the tensor is asymmetric. %An alternative method relies on the acquisition of RIXS spectra in several different crystallographic orientations while moving in reciprocal space, which provides sufficient data points to reconstruct the scattering tensor.

\begin{figure}
\includegraphics[width =  \columnwidth]{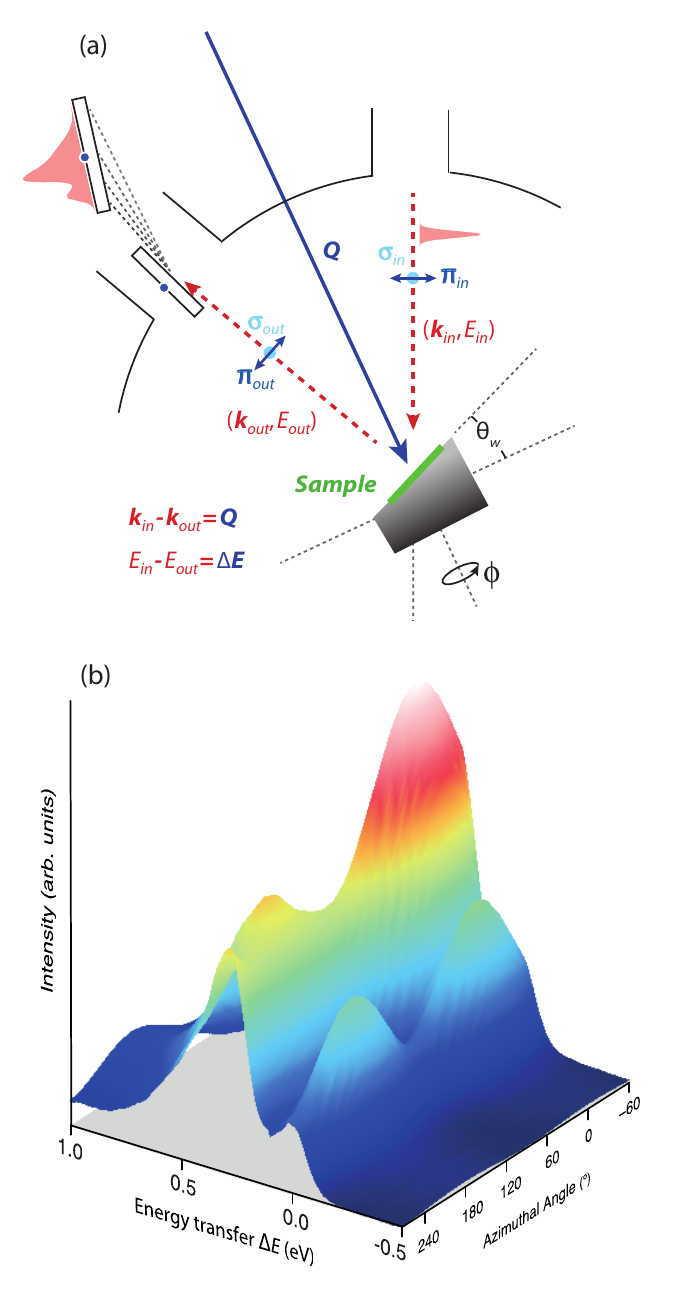}
\caption{\label{fig1} (a) Schematic representation of the scattering geometry for azimuthal angle dependent RIXS experiments. The orientation of the sample as in the figure defines the zero of the azimuthal angle $\phi$. (b) Surface plot of the azimuthal dependence of the low energy RIXS spectrum of AF-NCO measured with incoming $\sigma$ polarization.}
\end{figure}

%%%%%%%%%%%%%%%%%%%%%%%%%Fig1
In this paper, we apply a special procedure to resolve the RIXS scattering tensor \textit{at a given momentum transfer}, and correspondingly uncover the nature of excitations as a function of both energy and momentum. Our experimental approach relies on the use of an azimuthal scanning geometry where the sample is placed on a wedged holder as shown in Fig. 1(a). This geometry, owing to the collinearity of the azimuthal rotation axis to the direction of momentum transfer, ensures that the probed wavevector remains fixed (both in-plane $\mathbf{Q}_{\parallel}$ and out-of-plane $\mathbf{Q}_{\perp}$  components) for all values of the azimuthal angle ($\phi$). At each azimuthal angle, a different combination of the tensor components are selected, so that the symmetry of scattering tensor is imprinted onto the azimuthal dependence of RIXS signal. This probing scheme is often used in resonant elastic X-ray scattering experiments and is here demonstrated for inelastic processes \cite{comin_symmetry_2015,achkar_orbital_2014}. In our experiment, an ostensible variation of the RIXS signal can be observed as a function of $\phi$ and, most importantly, the intensity modulation is different for each spectral component, reflecting the symmetry of the underlying scattering tensor (Fig. 1(b)). We applied this method to resolve the charge, spin, and orbital nature of elastic scattering, magnon and multimagnon, and \textit{dd} excitations in cuprate compounds Nd$_{2}$CuO$_{4}$ (NCO) and YBa$_{2}$Cu$_{3}$O$_{6.75}$ (YBCO). The excellent agreement between theory and experiment for a wide range of excitations confirms the ability of our method to accurately disentangle the charge, spin, and orbital contributions to elementary excitations in solids.

%%%%%%%%%%%%%%%%%%%%%%%%%Sec2
\section{II. methods}

Thin films of antiferromagnetic and superconducting Nd$_{2}$CuO$_{4}$ (AF-NCO and SC-NCO) were grown by molecular beam epitaxy under ultra-high vacuum using Nd and Cu metal sources and atomic oxygen generated in-situ from a RF oxygen source \cite{krockenberger_emerging_2013}. Reflection high energy electron diffraction (RHEED) and electron impact emission spectroscopy (EIES) were used to monitor and control the growth of Nd$_{2}$CuO$_{4}$ films on (001) SrTiO$_{3}$ substrates in real time. High-resolution reciprocal space mapping data clearly show that the NCO films (100 nm) are grown fully relaxed. As-grown NCO films (AF-NCO) were rapidly cooled after the growth under ultra-high vacuum whereas SC-NCO were subject to a two-step annealing process \cite{krockenberger_unconventional_2012}. Superconductivity was confirmed by electric transport and magnetization measurements. The YBCO single crystals were grown by a self-flux method using BaZrO$_{3}$ crucibles \cite{liang_evaluation_2006}.

The RIXS data were collected at the ADRESS beamline of the Swiss Light Source at the Paul Scherrer Institute, Villigen PSI, Switzerland \cite{strocov_high-resolution_2010, ghiringhelli_saxes_2006}. The combined energy resolution was approximately 130 meV (determined by recording the elastic scattering from a carbon-filled acrylic tape), and the scattering angle was fixed to 130 degrees. The samples were mounted on a wedged-sample holder to align the azimuthal rotation axis with the direction of the probed wavevector $\textbf{Q}$. The azimuthal angle $\phi=0^{\circ}$ is defined so as to have the \textit{a} and \textit{c} crystallographic axes spanning the scattering plane. The azimuthal series ranges from $-60^{\circ}$ to $270^{\circ}$ in $15^{\circ}$ step, covering more than 90\% of the full $360^{\circ}$ rotation. All measurements were performed at 15 K using a liquid He cryostat. The excitation energy for RIXS measurements has been set to the maximum of the Cu $L_{3}$ absorption line and has been regularly monitored during the experiment. The data reported here have been normalized to the acquisition time as the intensity of the incoming beam was determined to be constant from monitoring the drain current on the last optical elements before the sample.

\begin{figure}
\includegraphics[width =  \columnwidth]{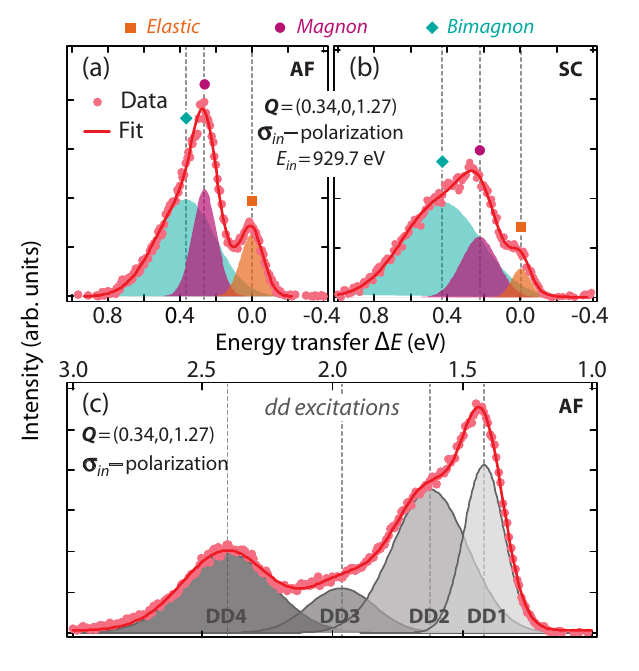}
\caption{\label{fig2} (a),(b) Low energy RIXS spectra of AF-NCO and SC-NCO after subtraction of a linear background. Orange, magenta, and green Gaussian peaks represent the elastic, (para-)magnon, and bimagnon peaks, respectively. (c) $\textit{dd}$ excitations region of AF-NCO. Data are measured at $\phi=0^{\circ}$ with incoming $\sigma$ polarization.}
\end{figure}

\begin{figure}
\includegraphics[width =  \columnwidth]{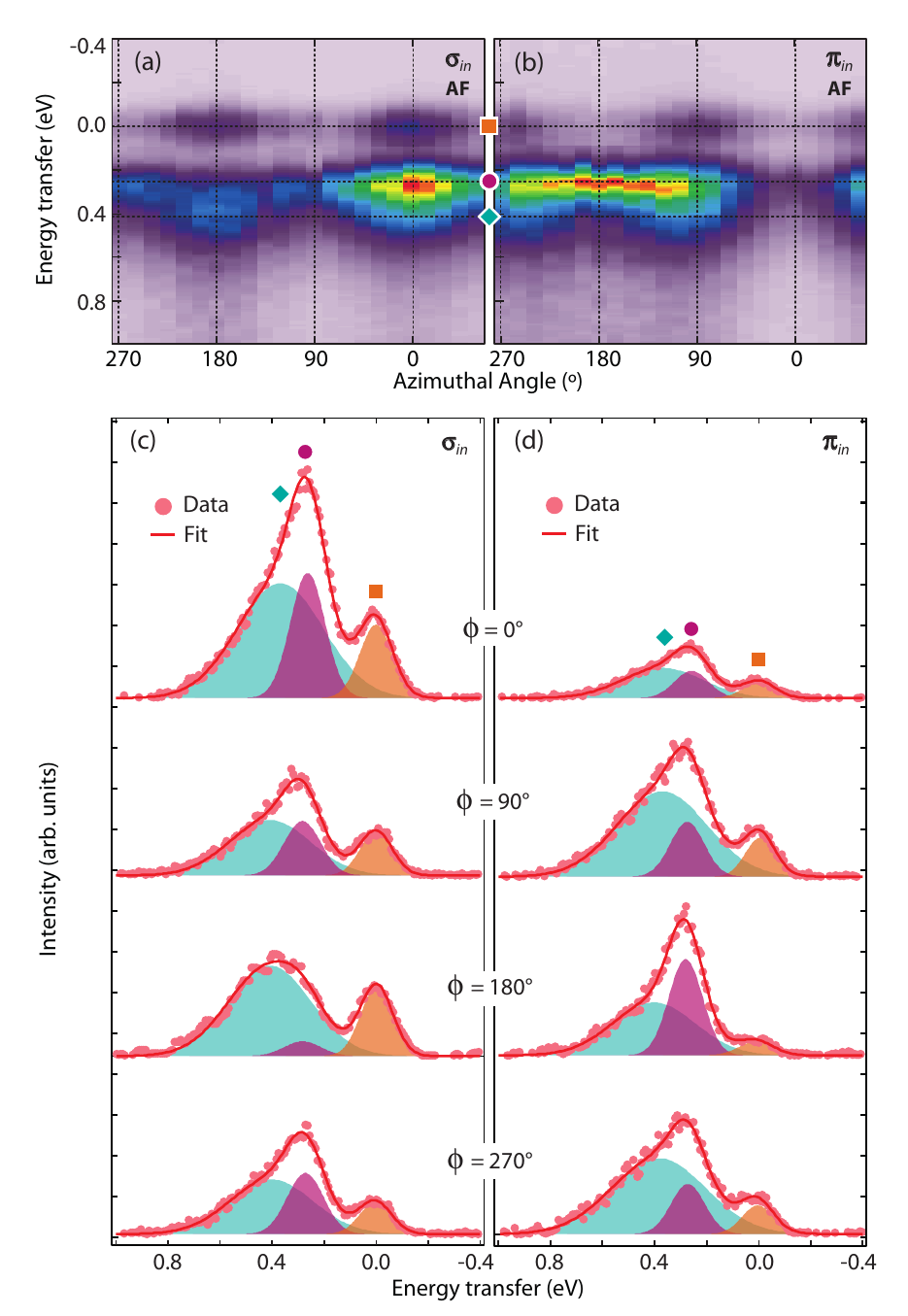}
\caption{\label{fig3} (a),(b) Azimuthal dependence of low energy RIXS spectra of AF-NCO measured with $\sigma$ and $\pi$ incoming polarization at \textbf{Q}=(0.34,0,1.27) $r.l.u.$. Positions of elastic, magnon, and bimagnon peaks at $\phi=0^{\circ}$, $90^{\circ}$, $180^{\circ}$, and $270^{\circ}$ are marked using orange, magenta, and green markers for guidance. (c),(d) Fitting of RIXS spectra at four representative azimuthal angles: $\phi=0^{\circ}$, $90^{\circ}$, $180^{\circ}$, and $270^{\circ}$.}
\end{figure}

%%%%%%%%%%%%%%%%%%%%%%%%%Sec3

\section{III. RIXS spectra of T$^{\prime}$-cuprates}

%%%%%%%%%%%%%%%%%%%%%%%%%Fig2

Figure 2(a) shows a representative Cu-$L_{3}$ RIXS spectrum of AF-NCO measured with $\sigma_{in}$ polarization, $\phi=0^{\circ}$, and \textbf{Q}=(0.34, 0, 1.27) reciprocal lattice units ($r.l.u$). As it is has been experimentally observed \cite{tacon_intense_2011,dean_persistence_2013,lee_asymmetry_2014,ishii_high-energy_2014,minola_collective_2015,braicovich_momentum_2010}, the low-energy RIXS spectrum of cuprates consists of elastic, magnon, and bimagnon (or multimagnon) peaks, which in AF-NCO are respectively located at $\Delta E=0$, $0.27\pm0.02$, $0.39\pm0.02$ eV ($\Delta E$ being the energy transfer: $E_{in} - E_{out}$).  Note that the energy of the bimagnon peak observed here is consistent with a bimagnon energy of 0.38 eV and 0.43 eV measured by optical Raman scattering and Cu-$K$ edge RIXS in the closely related compound La$_{2}$CuO$_{4}$ \cite{hill_observation_2008,lyons_dynamics_1988}. As illustrated in Fig. 2(b), low energy excitations in SC-NCO remain similarly well-resolved in spite of the (para-)magnon and bimagnon peaks becoming broader due to a doping-induced breakdown of long-range magnetic order \cite{tacon_intense_2011,dean_persistence_2013,lee_asymmetry_2014,ishii_high-energy_2014}. In Fig. 2(c), we also show the \textit{dd} excitation peaks from AF-NCO, which exhibit orbital components that appear well separated compared to other cuprates \cite{moretti_sala_energy_2011}. Considering the local $D_{4h}$ symmetry of the Cu$^{2+}$ ion, we assign the peaks at $\Delta E=1.42$, $1.67$, $2.40$ eV (DD1, DD2, and DD4 in Fig. 2(c)) to $d_{xy}$, $d_{xz/yz}$, and $d_{3z^{2}-r^{2}}$ orbital excitations, respectively. The additional, small peak near $1.95$ eV (DD3) has been observed in other $T^{\prime}$-cuprates such as CaCuO$_{2}$ and Sr$_{2}$CuO$_{2}$Cl$_{2}$, but the origin of this additional excitation is still under debate \cite{moretti_sala_energy_2011}.
%and has been suggested to originate from random oxygen vacancies that affect the local electronic structure of Cu$^{2+}$.

Before discussing the detailed azimuthal dependence of RIXS spectra and the comparison between experimental data and theoretical calculations, we remark that the self-absorption of scattered photons has to be accounted within the quantitative analysis of the X-ray scattering intensities in different geometries \cite{achkar_bulk_2011}. In our case, the measured atomic form factor allows us to calculate the effect of self-absorption as a function of experimental geometry and energy transfer, as in Ref. \cite{comin_symmetry_2015,achkar_orbital_2014}. In the Appendix C, we present the detailed procedure for self-absorption correction that was applied to our RIXS spectra. At the same time, we emphasize that the intensity variation in our experiment is clearly distinctive across different excitations and polarizations even before the self-absorption corrections (Appendix D). Thus, we can firmly rule out that self-absorption effects, which depend mainly on the experimental geometry, might hinder the reliability of our analysis \cite{achkar_bulk_2011}. All the data presented below are corrected for self-absorption accordingly.

\begin{figure*}[t]
\centering
\includegraphics[width =2\columnwidth]{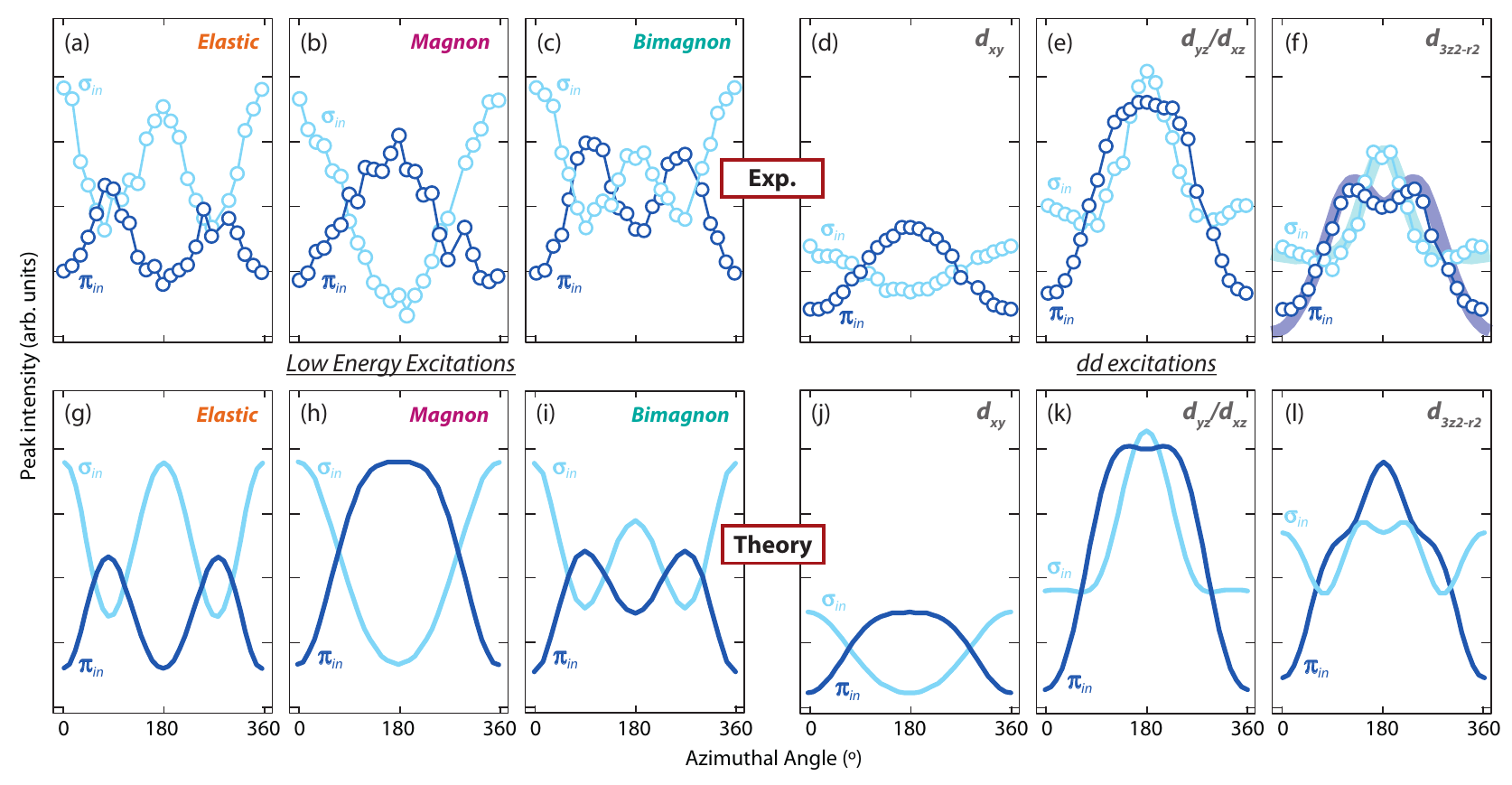}
\caption{\label{fig4} (a)-(c), Azimuthal dependence of low energy excitations of AF-NCO. Light blue (dark blue) open circles represent data measured with $\sigma$ ($\pi$) polarization. (d)-(f) Azimuthal dependence of \textit{dd} excitations of AF-NCO. (g)-(l), Corresponding calculations of azimuthal angle dependent RIXS intensity under the single ion approximation. The thick lines in (f) shows the calculation of azimuthal dependence of $d_{3z^{2}-r^2}$ excitation for spin-flip channel only (see text for details).}
\end{figure*}

\section{IV. Azimuthal dependence of the low-energy RIXS spectra}

%%%%%%%%%%%%%%%%%%%%%%%%%Fig3

The applicability of our approach is apparent from the pronounced azimuthal variation of the individual excitations in the RIXS spectra. This can be clearly appreciated from Fig. 3(a,b), which showcases the azimuthal dependence of the low-energy RIXS spectra of AF-NCO measured with $\sigma_{in}$ and $\pi_{in}$ polarized incoming light, respectively. Figure 3(c,d) displays a sample of individual spectra at four representative azimuthal angles $0^{\circ}$, $90^{\circ}$, $180^{\circ}$, and $270^{\circ}$. In the case of $\sigma_{in}$ incoming  polarization, it is clear from both the two-dimensional map, Fig. 3(a,b), and the peak decomposition of individual spectra, Fig. 3(c,d), that the cross-section for elastic scattering is maximized at $0^{\circ}$ and $180^{\circ}$, and minimized near $90^{\circ}$ and $270^{\circ}$. In stark contrast, the magnon intensity shows a maximum at $0^{\circ}$ and a minimum at $180^{\circ}$, with a doubled azimuthal periodicity compared to the elastic signal. With $\pi_{in}$ polarization, the position of maxima and minima is reversed, while the periodicity for each excitation is preserved.

%%%%%%%%%%%%%%%%%%%%%%%%%Fig4
In Fig. 4(a-c), we summarize the azimuthal dependence of low energy excitations of AF-NCO. For comparison, in Fig. 4(g),(h), we simulate the azimuthal dependence of elastic scattering and magnon excitations based on the scattering tensors obtained within the single-ion approximation (see Appendices for the details of calculations). Overall, the distinctive azimuthal dependences observed for different excitations, are well-reproduced by the calculation for both polarizations.

The scattering tensors for elastic scattering (non-spin-flip) and magnon excitations (single spin-flip) have the following form in the cuprates:
\begin{displaymath}
\mathit{F_{elastic}}=\left( \begin{array}{ccc}
2 & 0 & 0 \\
0 & 2 & 0 \\
0 & 0 & 0 \\
\end{array} \right), 
\mathit{F_{magnon}}=\left( \begin{array}{ccc}
0 & i & 0 \\
-i & 0 & 0 \\
0 & 0 & 0 \\
\end{array} \right)
\end{displaymath}
As expected, $F_{elastic}$ is reduced to a diagonal tensor since there is no transfer of photon angular momentum to the sample for elastic scattering. In contrast, magnetic excitations require transfer of the photon angular momentum to the spin degree of freedom via core-hole spin-orbit coupling. Thus, $F_{magnon}$ has nonzero off-diagonal components only \cite{ament_theoretical_2009}. We note that our form of $F_{magnon}$ in the single-ion approximation agrees with the general spin excitation tensor in the cuprates \cite{haverkort_theory_2010}.

An intuitive way to understand how different realizations of the scattering tensor are imprinted in the azimuthal dependence of the RIXS intensity is to consider the symmetries of the RIXS process. In the case of pure charge excitations, the intensity is dominated by polarization preserving ($\sigma\to\sigma^{\prime}$ and $\pi\to\pi^{\prime}$) channels. Thus, for most scattering events, the mirror symmetry in the scattering plane with respect to the scattering wave vector \textbf{Q} is preserved. The resulting symmetry between $I(\phi)$ and $I(180^{\circ}-\phi)$ forces the occurrence of two minima and two maxima over the full azimuthal range ($0^{\circ}$ to $360^{\circ}$). In contrast, for excitations with a single spin-flip, the intensity is dominated by polarization flipping ($\sigma\to\pi^{\prime}$ and $\pi\to\sigma^{\prime}$) channels which break the mirror symmetry with respect to \textbf{Q}. This gives rise to doubled periodicity of the intensity of (para-)magnon excitations compared to purely elastic scattering (charge) as clearly observed in our experiment.
 
The azimuthal dependence of the bimagnon excitation is expected to exhibit a mixed charge-like and spin-like character. In fact, as shown in Fig. 4(c), the azimuthal dependence of the bimagnon intensity displays two minima and maxima but with a clear asymmetry between $\phi^{\circ}=0$ and $\phi^{\circ}=180$ for both polarizations. This behavior can be reproduced by a superposition of charge and spin scattering as depicted in Fig. 4(i). This fitting of the azimuthal dependence allows us to reliably estimate the relative contribution of charge and spin spectral weight to be 2.7$\pm0.2$. The dominance of charge-like ($\Delta S=0$) processes in the bimagnon spectral weight naturally explains the resemblance of bimagnon spectra of La$_{2}$CuO$_{4}$ measured at the Cu-$L_{3}$ and Cu-$K$ edge, where in the latter case only $\Delta S=0$ bimagnon excitations are allowed \cite{bisogni_bimagnon_2012}.

%%%%%%%%%%%%%%%%%%%%%%%%%Fig5

\begin{figure*}[t]
\centering
\includegraphics[width =2\columnwidth]{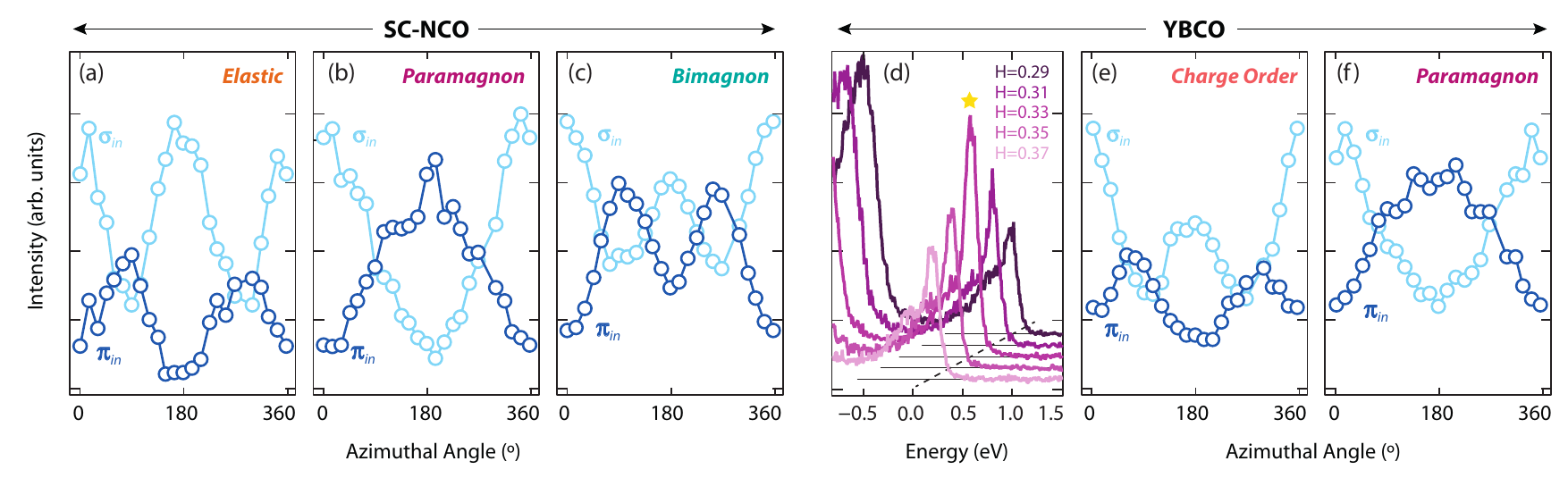}
\caption{\label{fig5} (a)-(c) Azimuthal dependence of elastic scattering, paramagnon excitation, and bimagnon excitation in SC-NCO. (d) Low energy excitations of YBCO around the charge order wave vector. (e),(f) Azimuthal dependence of the charge order scattering and the paramagnon excitation in YBCO.}
\end{figure*}

In Fig. 5, we extended our analysis to the doped cuprates, SC-NCO and YBCO. In case of SC-NCO, the azimuthal dependence of elastic scattering, paramagnon excitation, and bimagnon excitation excellently agrees with the result from AF-NCO. Our results strongly support the hypothesis that the (para-)magnon and bimagnon peaks in SC-NCO possess the same character as in the parent insulator. At the same time, fitting to azimuthal dependence of bimagnon excitation in SC-NCO reveals that charge-to-spin ratio slightly increased with doping (from 2.7$\pm0.2$ to 3.0$\pm0.2$), indicating the additional contribution to charge spectral weight from doped carriers (see also section VI) \cite{benjamin_single-band_2014,kanasz-nagy_resonant_2016,guarise_anisotropic_2014}.

In the case of YBCO, we used the experimental geometry that locks the in-plane momentum transfer ($\mathbf{Q_{\parallel}}$) at the charge order wave vector \textit{h}$\sim$0.33 \textit{r.l.u.}, where the RIXS spectrum is characterized by a pronounced enhancement of the elastic line as illustrated in Fig. 5(d) \cite{ghiringhelli_long_2012}. The azimuthal dependence of the elastic scattering intensity of YBCO shows two minima/maxima as expected for charge scattering, though the intensity is partly suppressed near $\phi=180^{\circ}$ due to a slight misalignment between in-plane momentum transfer and the exact charge order wave vector \cite{footnote}. On the other hand, the azimuthal dependence of the paramagnon intensity in YBCO quantitatively agrees with that of AF-NCO and SC-NCO. In conclusion, the characteristic azimuthal dependence of charge and spin excitations shows high consistency across all samples investigated, confirming the high reliability of our methods.

%%%%%%%%%%%%%%%%%%%

\section{V. Azimuthal dependence of the orbital excitations}

With the sensitivity to the detailed azimuthal dependence of low energy excitations, our method can be further applied to resolve the excitations with more complex scattering tensor. In cuprates, this is the case for the \textit{dd} excitations at higher energy, where the scattering matrices are intrinsically asymmetric and contain a larger number of non-vanishing matrix elements: For example, the scattering matrix for non-spin flip excitation from $d_{x^2-y^2}$ to $d_{xz}$ is

\begin{displaymath}
\mathit{F_{d_{x^{2}-y^{2}}^{\uparrow}\rightarrow d_{xz}^{\uparrow}}}=\left( \begin{array}{ccc}
\frac{1}{2} & \frac{i}{2} & 0 \\
-\frac{i}{2} & -\frac{1}{2} & 0 \\
\frac{1}{2} & -\frac{3i}{2} & 0 \\
\end{array} \right)
\end{displaymath}

The pronounced asymmetry is due to the different orbital quantum numbers between initial and final states, which forces the RIXS process to occur via transfer of photon angular momentum to orbital angular momentum. This results in the complex azimuthal dependence of RIXS intensity as calculated in Fig. 4(j-l). Through this analysis, we can assign the experimentally observed azimuthal dependence of $d_{xy}$ and $d_{yz}/d_{xz}$ excitations to corresponding calculations for different orbital excitations, as displayed in Fig. 4(d,e). The excellent agreement demonstrates that our approach is not limited to the charge and spin excitations, and is applicable in resolving the excitations involving complex and very general (or low-symmetric) form of the scattering tensor.

It is worth noting that the azimuthal dependence of the $d_{3z^2-r^2}$ excitation shows an ostensible discrepancy with the calculation. The deviation from the calculation is manifested differently for the $\sigma_{in}$ and $\pi_{in}$ polarization channels (for example, the experimental intensity at $\phi$=180$^{\circ}$ appears to be higher than the calculation in $\sigma_{in}$ channel, while it is lower in $\pi_{in}$ channel), so we rule out any geometrical causes (for example, inaccuracies in self-absorption correction). Rather, this unexpected azimuthal dependence might reflect an intrinsic property of the sample. Surprisingly, the experimental azimuthal dependence can be perfectly reproduced if we only account the $d_{3z^2-r^2}$ excitation with spin-flip, shown as a thick line in Fig. 4(f). This implies that the spin-conserving $d_{x^{2}-y^{2}}\rightarrow d_{3z^2-r^2}$ excitation is intrinsically suppressed in the CuO plane. The origin of this unexpected behavior of $d_{3z^2-r^2}$ orbital excitation in $T^{\prime}$-cuprate might be beyond the single-ion approximation, and will be the subject of future theoretical studies.

We also investigated the azimuthal dependence of the additional peak observed in the $dd$ excitation region (DD3 in Fig. 2(c)) to shed new light on its origin. This additional peak has also been observed in other $T^{\prime}$-structured cuprates, but its nature has remained unclear \cite{moretti_sala_energy_2011}. Previously, it has been suggested to originate from random oxygen vacancies that affect the local electronic structure of Cu$^{2+}$ \cite{moretti_sala_energy_2011}. However in our case, the peak appears in both as-grown AF-NCO and two-step oxygen-annealed SC-NCO with nearly equal intensity, suggesting that this excitation might not be related to random oxygen vacancies. Moreover, the peak has a systematic azimuthal dependence (Appendix E), implying that it arises from an excitation with a well-defined spin/orbital symmetry, rather than just random defects. Combined with the unexpected azimuthal dependence of $d_{3z^2-r^2}$ excitation, this shows that there might be rich but yet unexplored physics behind the $dd$ excitation part of the RIXS spectrum of $T^{\prime}$-cuprates.

%The only major difference occurs near for the $d_{3z^{2}-r^{2}}$ orbital excitation, which might be due to the hybridization with interstitial oxygen atoms or interlayer interactions not captured in the single ion approximation. 

Altogether, the broad agreement between experiment and theory for charge, spin, and orbital excitations demonstrates that azimuthal angle dependent RIXS can be used to unfold the contribution of multiple elements of the scattering matrix. The proposed approach can be used to identify the character of local and collective excitations in several materials with coupled spin/charge/orbital degrees of freedom such as cobaltates, Fe-based superconductors, orbital-ordered maganites, and spin-orbit coupled 5$d$ oxides \cite{wilkins_direct_2003, zhou_persistent_2013,pelliciari_intralayer_2016,pelliciari_local_2017,satoh_excitation_2017,kim_magnetic_2012,calder_spin-orbit-driven_2016}.

%%%%%%%%%%%%%%%%%%%%%%%%%Fig6

\section{VI. Disentangling the RIXS spectra in terms of the nature of excitations}

\begin{figure}
\includegraphics[width =  \columnwidth]{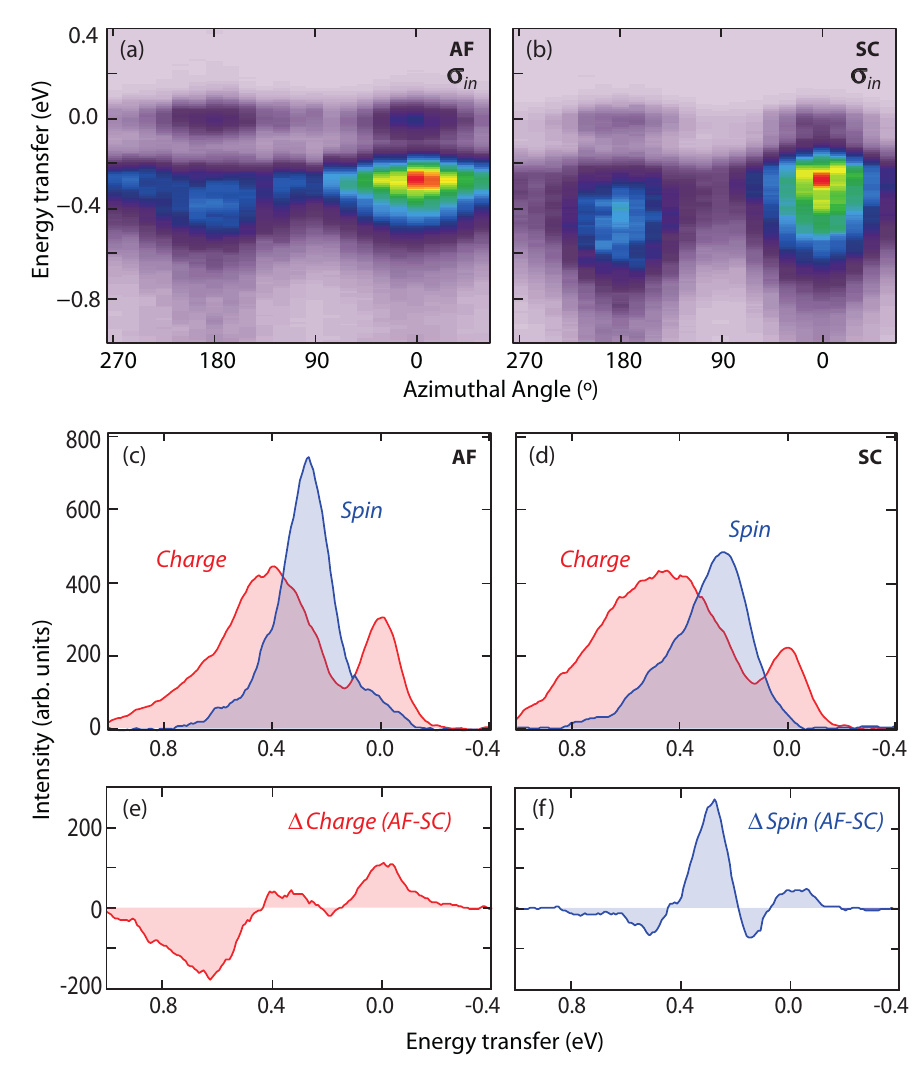}
\caption{\label{fig6} (a),(b) Azimuthal dependence of RIXS spectra of AF-NCO and SC-NCO measured with $\sigma$ incoming polarization at \textbf{Q}=(0.34, 0, 1.27) $r.l.u.$ (c),(d) Disentangled intensity of charge-like and spin-like excitations in RIXS spectra of AF-NCO and SC-NCO. (e),(f) Difference of the spin and charge spectral distribution of AF-NCO and SC-NCO.}
\end{figure}

In the last part of our analysis, we extend the machinery developed thus far to extract the charge- ($\Delta S=0$) and spin-like ($\Delta S=1$) components of the RIXS spectrum. Figure 6(a,b) report the azimuthal dependence of the RIXS spectrum of AF-NCO and SC-NCO measured with incident $\sigma$ polarized light. Using these $(E, \phi)$ RIXS maps, we can separate the charge and spin contributions from the RIXS azimuthal dependence (at every value of energy transfer) using the following equation:
\begin{equation}
I(E,\phi)=w_{c}(E)\cdot I_{c}(\phi)+w_{s}(E)\cdot I_{s}(\phi)
\end{equation}
Here $I_{c(s)}(\phi)$ is the calculated azimuthal dependence of the charge (spin) excitation intensity (Fig. 4(g,h)) and $w_{c(s)}(E)$ is the fitting parameter representing respective weight of charge (spin) at each energy transfer. Considering the similarity of $I_{c(s)}(\phi)$ with and $\cos2\phi$ ($\cos\phi$) function, this procedure is similar to fourier transform in $\phi$ domain where $\pi$ ($2\pi$) periodic component represents the weight from charge (spin) channel. As displayed in Fig. 6(c,d) the spin and charge components of the RIXS intensity can be effectively extracted using this method. The charge-like contribution dominates the intensity of the elastic scattering and bimagnon excitation, while spin excitations sharply peak at the single magnon/paramagnon energy. We note that the former and the latter show strong resemblance to the RIXS spectra of Ce-doped bulk NCO measured in $\sigma\rightarrow\sigma$ and $\sigma\rightarrow\pi$ polarization channels respectively \cite{Neto_RIXSNCCO}. Additionally, our approach accurately captures subtle features such as an asymmetry in the spin spectral distribution (SD) near 400 meV due to spin contributions within the bimagnon excitations, and the broadening of magnetic excitations in SC-NCO due to the damping by the Stoner excitation \cite{tacon_intense_2011,dean_persistence_2013,lee_asymmetry_2014,ishii_high-energy_2014}. This finding underscores the high sensitivity of azimuthal angle dependent RIXS in resolving the character of electronic excitations, even in presence of substantial spectral overlap.

The present method enables to separately track down the evolution of the spectral distribution from charge- and spin-like excitations as a function of doping. This information is pivotal to understand the role of low-energy excitations in the framework of high temperature superconductivity. The mixed spin-charge character of the bimagnon peak also indicates that the spin and charge channels cannot be simply separated by isolating the magnon and bimagnon features, but require a more detailed analysis of the kind discussed in this study. In Fig. 6(e,f), we present the SD difference between AF-NCO and SC-NCO ($\Delta SD=SD_{AF}-SD_{SC}$) in the charge and spin channels. The main finding is that the spin and charge spectral distributions are unequally affected by doping: In the superconducting phase, the total spin spectral weight decreases by $\approx$10 \%, while the charge spectral weight increases by a similar amount. The difference of spectral distribution in the spin channel displays a positive peak at the single (para-)magnon energy and two side minima, which likely arise from the broadening of both (para-)magnon and spin-like bimagnon excitations upon doping. Except for the slight broadening, the shape and weight of spin SD of AF-NCO largely survives in SC-NCO, consistent with the persistence of spin-fluctuations in the superconducting regime \cite{tacon_intense_2011,dean_persistence_2013,lee_asymmetry_2014,ishii_high-energy_2014}. In contrast, we observe additional spectral weight in the charge channel of SC-NCO at high energy (from 0.5 eV to 1 eV), which cannot be attributed to the broadening of charge-like bimagnon excitations ($\Delta S=0$). Instead, this spectral weight might originate from additional contributions due to particle-hole excitations in the doped compound, as suggested in recent theoretical studies \cite{benjamin_single-band_2014,kanasz-nagy_resonant_2016}. The energy scale of this particle-hole contribution is similar to that of the broad continuum in overdoped YBCO which exhibits fluorescence behavior \cite{minola_collective_2015}.
%Our observation of a distinct evolution of spin and charge spectral distribution with doping casts new light on the physics of electron- as well as hole-doped copper-oxide superconductors and is expected to spur new investigations on the evolution of the spin and charge spectral intensity in these systems.

%%%%%%%%%%%%%%%%%%%%%%%%%Sec5

\section{VII. Summary}

In summary, we have demonstrated a systematic method to resolve the charge, spin, and orbital character of the electronic excitations in RIXS spectra. In the study of prototypical cuprate compounds, we observed distinct azimuthal dependences of elastic scattering, (para-)magnon/bimagnon excitations, and \textit{dd} excitations, and find them in excellent agreement with theoretical calculations within the single-ion approximation. Ultimately, our method has been used to disentangle the spin and charge contribution to the low-energy RIXS spectra of cuprates. The ability to track down the evolution of the charge- and spin-like character of excitations from the antiferromagnetic to the superconducting regime, reveals important new information on the excitations that are more strongly coupled to the electronic states in the CuO planes. The method presented in this work introduces a new way to use RIXS to resolve the character of electronic excitations in a wide range of quantum materials with entangled charge, spin, and orbital degrees of freedom.

\section{Acknowledgement}

We thank Senthil Todadri and Krzysztof Wohlfeld for insightful discussions. We acknowledge Grace H. Zhang for supporting calculations. We thank Leonard Nue for help in manufacturing the wedged-sample holders. We acknowledge Paul Scherrer Institut for provision of synchrotron radiation beamtime at the ADRESS beamline of the Swiss Light Source. M.K. acknowledges a Samsung Scholarship from the Samsung Foundation of Culture. J.P. is financially supported by the Swiss National Science Foundation Early Postdoc, and Postdoc Mobility fellowship Project No. P2FRP2\_171824 and P400P2\_180744 E.P. is financially supported by the Swiss National Science Foundation (SNSF) through the Sinergia network Mott Physics Beyond the Heisenberg (MPBH) model, and D.E.M. is supported by the NCCR MARVEL of SNSF.

%%%%%%%%%%%%%%%%%%%%%%%%APPEN

\section{Appendix A: Scattering matrices OF CHARGE/SPIN/ORBITAL EXCITATIONS in RIXS}

In resonant inelastic x-ray scattering (RIXS), the intensity of the scattered electric field can be expressed as:
\begin{eqnarray}
I(\omega_{k},\omega_{k}^{\prime},\mathbf{k},\mathbf{k}^{\prime},\epsilon,\epsilon^{\prime})&=&\sum_{f}|F(\omega_{k},\omega_{k}^{\prime},\mathbf{k},\mathbf{k}^{\prime},\epsilon,\epsilon^{\prime})|\nonumber\\
&\times&\delta(E_{f}+\hbar\omega_{k^{\prime}}-E_{g}-\hbar\omega_{k})
\end{eqnarray}
where $\omega_{k},\mathbf{k},\epsilon$ ($\omega_{k^{\prime}},\mathbf{k}^{\prime},\epsilon^{\prime}$) represent the energy, momentum and polarization of the incident (outgoing) light, and $F$ is the RIXS scattering amplitude \cite{ament_resonant_2011}. The summation runs over all possible final states $f$, and the $\delta$-function enforces the energy conservation explicitly.

In $T^{\prime}$-cuprates, the Cu$^{2+}$ ion is surrounded by four oxygen ions and has $D_{4h}$ point group symmetry. Under the single ion approximation and the dipole approximation, the RIXS intensity for each final state at Cu $L_{3}$  edge is proportional to the following matrix elements which in the second order perturbation expansion reads:
\begin{eqnarray}
I_{f}^{RIXS}(\epsilon,\epsilon^{\prime})&\propto&|\sum_{i}\langle f|D_{\epsilon^{\prime}}^{\dag}|i\rangle\langle i|D_{\epsilon}|g\rangle|^{2}\nonumber\\
&=&|\sum_{m}\langle f|D_{\epsilon^{\prime}}^{\dag}|2p_{3/2,m}\rangle\langle 2p_{3/2,m}|D_{\epsilon}|g\rangle|^{2}
\end{eqnarray}
where $|g\rangle$, $|i\rangle$, and $|f\rangle$ represent the ground state, intermediate state with core-hole, and final state respectively, and $D_{\epsilon}$ is a dipole operator. In cuprates, Cu spins are antiferromagnetically aligned along the Cu-O bond direction \cite{RMPecuprates}, therefore the state $d_{x^{2}-y^{2}}^{\uparrow}$ can be rewritten as $(d_{x^{2}-y^{2}}^{\uparrow z}+d_{x^{2}-y^{2}}^{\downarrow z})/\sqrt{2}$. Depending on the final state, the above equation describes the elastic scattering for $|f\rangle=3d_{x^{2}-y^{2}}^{\uparrow}$, the spin excitations for $|f\rangle=3d_{x^{2}-y^{2}}^{\downarrow}$, and the orbital+spin excitations for $|f\rangle=3d_{xy}, 3d_{xz}, 3d_{yz}, 3d_{3z^{2}-r^{2}}$. In the calculation of the intensity of orbital excitations, we summed over all possible final spin states since the spin splitting for the \textit{dd} excitation is not resolved in our experiment. Theoretically, this can be justified by the fact that the super-exchange coupling \textit{J} is orbital dependent, and negligible except for $d_{x^{2}-y^{2}}$ orbitals \cite{moretti_sala_energy_2011}.

Based on the above equation, we can express the scattering matrices in the sample frame. For elastic scattering and magnon excitations, the orbital symmetry is preserved, and the resulting matrix ($F_{elastic}$ and  $F_{magnon}$) has relatively simple form as described in Sec IV. The scattering tensor for orbital excitations is more complex and asymmetric due to the change of orbital symmetry before and after the scattering event. For example, the $d_{3z^{2}-r^{2}}$ orbital excitation matrices for all possible spin channels are given as:

\begin{displaymath}
\mathit{F_{d_{x^{2}-y^{2}}^{\uparrow}\rightarrow d_{3z^{2}-r^{2}}^{\uparrow}}}=\left( \begin{array}{ccc}
0 & -\frac{2i}{\sqrt{3}} & 0 \\
-\frac{2i}{\sqrt{3}} & 0 & 0 \\
0 & \frac{2i}{\sqrt{3}} & 0 \\
\end{array} \right)
\end{displaymath}
\begin{displaymath}
\mathit{F_{d_{x^{2}-y^{2}}^{\downarrow}\rightarrow d_{3z^{2}-r^{2}}^{\downarrow}}}=\left( \begin{array}{ccc}
0 & -\frac{2i}{\sqrt{3}} & 0 \\
-\frac{2i}{\sqrt{3}} & 0 & 0 \\
0 & \frac{2i}{\sqrt{3}} & 0 \\
\end{array} \right)
\end{displaymath}
\begin{displaymath}
\mathit{F_{d_{x^{2}-y^{2}}^{\uparrow}\rightarrow d_{3z^{2}-r^{2}}^{\downarrow}}}=\left( \begin{array}{ccc}
-\frac{1}{\sqrt{3}} & 0 & 0 \\
0 & \frac{1}{\sqrt{3}} & 0 \\
\frac{2}{\sqrt{3}} & 0 & 0 \\
\end{array} \right)
\end{displaymath}
\begin{displaymath}
\mathit{F_{d_{x^{2}-y^{2}}^{\downarrow}\rightarrow d_{3z^{2}-r^{2}}^{\uparrow}}}=\left( \begin{array}{ccc}
-\frac{1}{\sqrt{3}} & 0 & 0 \\
0 & \frac{1}{\sqrt{3}} & 0 \\
-\frac{2}{\sqrt{3}} & 0 & 0 \\
\end{array} \right)
\end{displaymath}

As expected, the matrix elements representing a positive transfer of angular momentum from the sample to the photon (below diagonal) are nonzero, while those representing a negative transfer (above diagonal) become vanishing. This complexity of the orbital excitation tensor leads to a nontrivial azimuthal dependence of the RIXS intensity as discussed in Sec V.

\begin{figure}
\includegraphics[width = 0.8 \columnwidth]{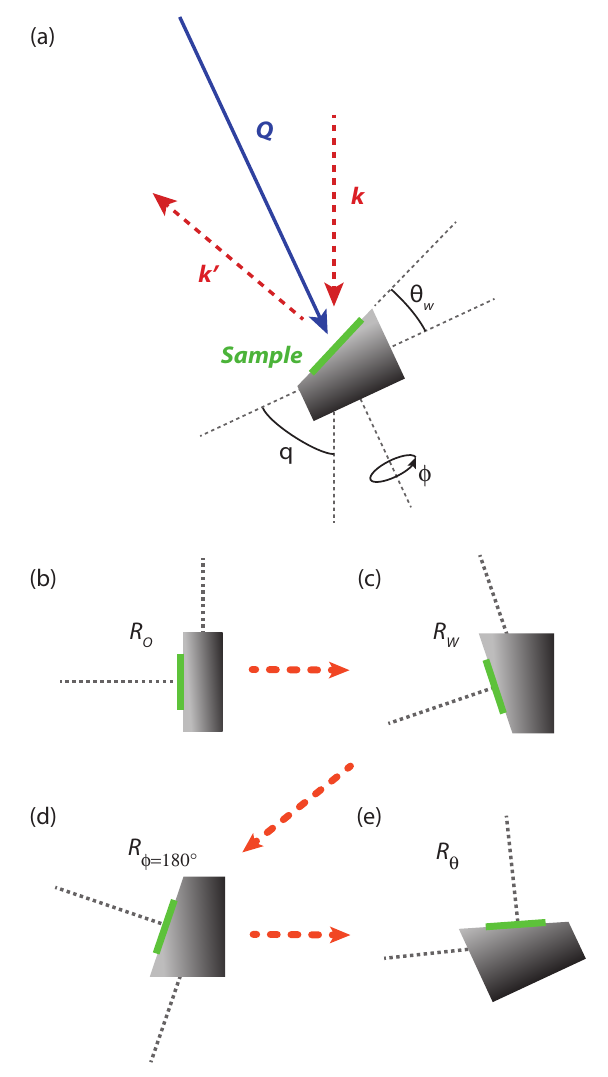}
\caption{\label{fig7} (a) Scattering geometry of our experiment and definition of the rotation angles. (b)-(e) Schematics for the sequential operations of $R_{o}$, $R_{w}$, $R_{\phi}$, and $R_{\theta}$.}
\end{figure}

\section{Appendix B: Calculation of the azimuthal dependence of RIXS intensity}

The above scattering matrices are defined in the sample frame, while the polarization vectors of incoming and outgoing photon are defined in the laboratory frame. Thus, we need an appropriate rotation matrix $R_{tot}(\theta_{w},\phi,\theta)$ before projecting the scattering matrix onto the polarization vectors.

The total rotation matrix $R_{tot}(\theta_{w},\phi,\theta)$ can be constructed from the four sequential rotation operations: $R_{o}$ rotates the sample frame to $\theta_{w}=\phi=\theta=0$ position in the laboratory frame; $R_{w}$ implements the transformation to the wedge sample holder configuration; $R_{\phi}$ is responsible for the azimuthal rotation; and $R_{\theta}$ aligns the axis of the azimuthal rotation with the scattering wave vector \textbf{Q}=$\mathbf{k}^{\prime}-\mathbf{k}$. Therefore we have $R_{tot}(\theta_{w},\phi,\theta)=R_{o}R_{w}R_{\phi}R_{\theta}$. The angles $\theta_{w}$, $\phi$, $\theta$ are defined in Fig. 7(a), and the schematics for each rotation is shown in Fig. 7(b)-(e). In our experiment, $\theta_{w}$ and $\theta$ are fixed to 40$^{\circ}$ and 65$^{\circ}$ respectively, leaving $\phi$ as the only variable of $R_{tot}$. After applying the full rotation operation to the scattering tensor corresponding to each type of excitation, we can directly calculate the azimuthal dependence of the RIXS intensity as:
\begin{equation}
I_{f}^{RIXS}(\epsilon,\phi)=\sum_{\epsilon^{\prime}}|\epsilon^{\prime}\cdot R_{tot}^{T}(\phi)F_{f}R_{tot}(\phi)\cdot\epsilon|^{2}
\end{equation}
Here, we sum the intensity over the polarization of the outgoing photons since the latter is not experimentally determined in the present experiment. The essence of azimuthal dependence analysis is that, via the action of the azimuthal angle dependent unitary matrix $R_{tot}(\phi)$, the individual elements of the scattering tensor can be disentangled once projected to fixed polarization vectors. In other words, the form of the scattering tensor and thus the nature of the corresponding excitation is imprinted onto the  dependence of $I^{RIXS}_{f}(\epsilon,\phi)$.\\

\section{Appendix C: Self-absorption correction in RIXS}

\begin{figure}
\includegraphics[width=\columnwidth]{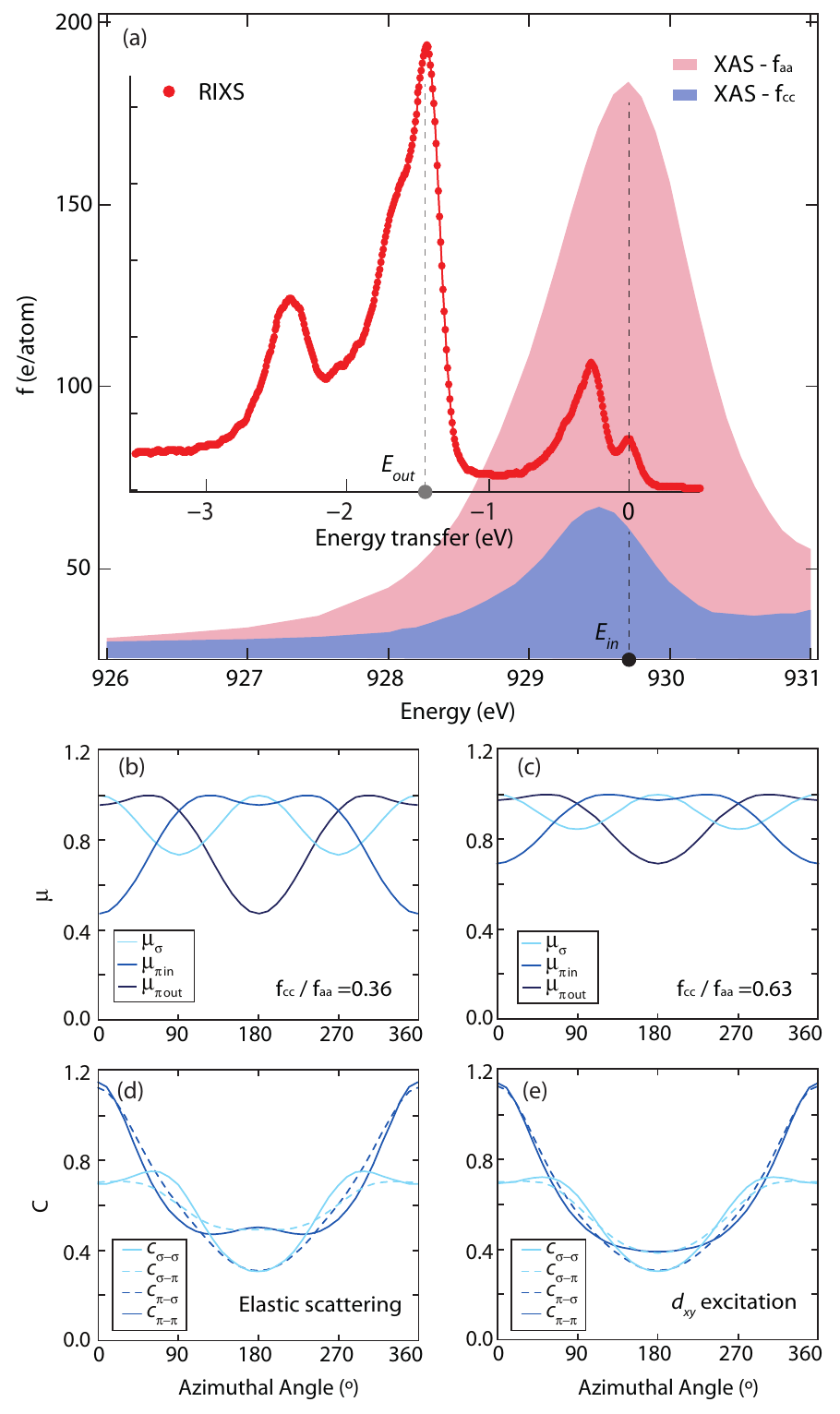}
\caption{\label{fig8} (a) XAS and RIXS spectrum of AF-NCO. (b),(c) Azimuthal dependence of the absorption coefficients for $f_{cc}$/$f_{aa}$=0.36 and $f_{cc}$/$f_{aa}$=0.63. (d),(e) Azimuthal dependence of the self-absorption correction factors for elastic scattering and $d_{xy}$ orbital excitation.}
\end{figure}

\begin{figure*}[t]
\centering
\includegraphics[width =2\columnwidth]{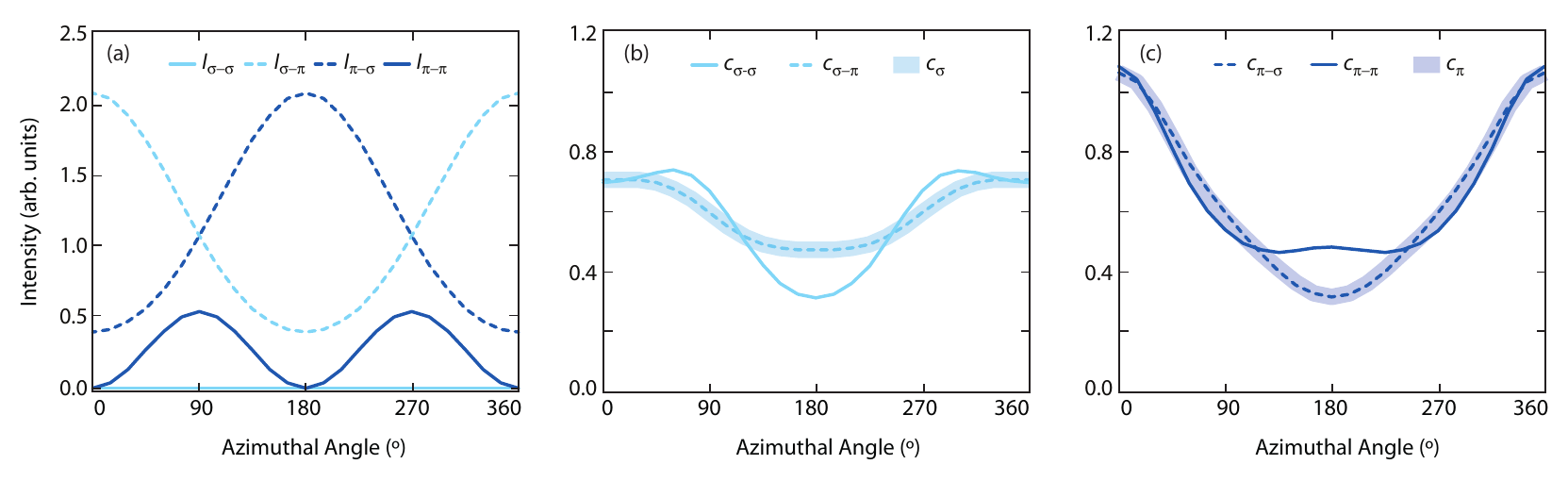}
\caption{\label{fig9} (a) Calculated (para-)magnon intensities for different polarization channels. (b),(c) Outgoing-polarization-resolved and total self-absorption correction factor for $\sigma$ and $\pi$ incoming polarization.}
\end{figure*}

Self-absorption is an unavoidable contribution to the measured intensity of scattered X-rays, and depends on the experimental geometry and energy loss of the photons at a given incident photon energy. Generally, the removal of self-absorption effects in the raw data is precluded in the absence of outgoing polarization-resolved data. However, in our case, the effect of self-absorption can be exactly calculated in the model, given a specific form of the scattering tensor \cite{comin_symmetry_2015,achkar_orbital_2014}. Specifically, the following formula is used to correct for self-absorption effects \cite{achkar_bulk_2011}:

\begin{eqnarray}
I_{corr}&=&I_{theory}\times[\mu_{in}(E_{in},\epsilon_{in},\phi)\nonumber\\
&+&\mu_{out}(E_{in},\epsilon_{in},\phi)\times\frac{-\hat k_{in}\cdot\hat n(\phi)}{\hat k_{out}\cdot\hat n(\phi)}]^{-1}\nonumber\\
&=&I_{theory}\times C(E_{in},\epsilon_{in},E_{out},\epsilon_{out},\phi)
\end{eqnarray}

Here, $\mu_{in}$ and $\mu_{out}$ are, respectively, the absorption coefficient for the incoming and outgoing light, and $\hat n$ is the vector normal to the sample surface. The dot product of $\hat n$ and $\hat k_{in}$ or  $\hat k_{out}$ defines the geometrical factor for self-absorption correction. The $\phi$ dependence of the correction function $C$ originates from the absorption coefficient as well as the geometrical factor.

The explicit dependence of the absorption coefficient on the experimental parameters can be written as:

\begin{eqnarray}
\mu_{in(out)}=\epsilon_{in(out)}\cdot[&R^{T}(\phi)&\nonumber\\
\left(\begin{array}{ccc}
f_{aa}(E) & 0 & 0 \\
0 & f_{aa}(E) & 0 \\
0 & 0 & f_{cc}(E) \\
\end{array} \right)&R(\phi)&]\cdot\epsilon_{in(out)}
\end{eqnarray}

In our experiment $E_{in}$ is fixed to the energy of the Cu $L_3$ edge (in our case, 929.7 eV), but the emission energy of the photons $E_{out}$, and thus $f_{aa}(E_{out})$ and $f_{cc}(E_{out})$, vary for different excitations. The energy-dependent values of $f_{aa}$ and $f_{cc}$ can be obtained from X-ray absorption spectra (total electron yield) collected at different polarization. In Fig 8(a), we show the absorption spectra of parent Nd$_{2}$CuO$_{4}$ (AF-NCO), overlapped with the excitation spectrum from RIXS. The $f_{cc}$/$f_{aa}$ ratio is minimum at the peak of the absorption spectrum, and approaches unity for higher energy transfers. For example, $f_{cc}$/$f_{aa}$=0.36 for elastic scattering, and $f_{cc}$/$f_{aa}$=0.63 for the $d_{xy}$ excitation.

Figure 8(b,c) show the dependence of the absorption coefficient on the polarization and azimuthal angle, for different values of  $f_{cc}$/$f_{aa}$ [corresponding to different excitations in the RIXS spectrum of Fig. 8(a)]. The variation with azimuthal angle is maximized for small  $f_{cc}$/$f_{aa}$, and is suppressed as  $f_{cc}$/$f_{aa}$ approaches unity for high energy excitations. The azimuthal dependence of $\mu$ possesses mirror symmetry with respect to $\phi=180^{\circ}$ due to the mirror symmetry relative to the scattering plane. Additionally, the profiles of $\mu_{\pi_{in}}$ and $\mu_{\pi_{out}}$ are shifted by $180^{\circ}$ due to mirror symmetry with respect to scattering wave vector Q.

Figure 8(d,e) display the calculated correction factor $C$ for elastic scattering and for $d_{xy}$ excitations. Overall, the effect of self-absorption correction is maximized for a grazing emission geometry ($\phi=180^{\circ}$), and minimized for a grazing incidence geometry ($\phi=0^{\circ}$), as expected. The outgoing polarization dependence of $C$ is suppressed at high $f_{cc}/f_{aa}$, and curves with different outgoing polarizations and eventually collapses onto a single curve when $f_{cc}/f_{aa}=1$.

To calculate the RIXS intensity for $\sigma$ and $\pi$ incoming polarization, the self-absorption correction should be applied separately to each polarization channel, as follows:

\begin{eqnarray}
I_{\sigma}^{corr}(\phi)=I_{\sigma\rightarrow\sigma}\times C_{\sigma\rightarrow\sigma}+I_{\sigma\rightarrow\pi}\times C_{\sigma\rightarrow\pi}\\
I_{\pi}^{corr}(\phi)=I_{\pi\rightarrow\sigma}\times C_{\pi\rightarrow\sigma}+I_{\pi\rightarrow\pi}\times C_{\pi\rightarrow\pi}
\end{eqnarray}

With the exact knowledge of $I_{\sigma(\pi)\rightarrow\sigma}$ and $I_{\sigma(\pi)\rightarrow\pi}$ from calculation (see Appendix A and B), we can combine $C_{\sigma(\pi)\rightarrow\sigma}$ and $C_{\sigma(\pi)\rightarrow\pi}$ to the single factor $C_{\sigma(\pi)}$, which depends only on the incoming polarization.

\begin{eqnarray}
I_{\sigma(\pi)}^{corr}&=&[I_{\sigma(\pi)\rightarrow\sigma}+I_{\sigma(\pi)\rightarrow\pi}]\\
&\times&\frac{I_{\sigma(\pi)\rightarrow\sigma}\times C_{\sigma(\pi)\rightarrow\sigma}+I_{\sigma(\pi)\rightarrow\pi}\times C_{\sigma(\pi)\rightarrow\pi}}{[I_{\sigma(\pi)\rightarrow\sigma}+I_{\sigma(\pi)\rightarrow\pi}]}\\
&=&I_{\sigma(\pi)}(\phi)\times C_{\sigma(\pi)}(\phi)
\end{eqnarray}

In Fig. 9, we report an example of the procedure outlined above. Figure 9(a) shows the calculated (para-)magnon intensity for each incoming and outgoing polarization. The RIXS intensity of (para-)magnon excitations is dominated by cross-polarization channels as expected. The intensity of $\sigma\rightarrow\sigma$ scattering vanishes for all azimuthal angles, as the scattering processes in this channel do not transfer photon anglular momentum to the sample. In contrast, the $\pi\rightarrow\pi$ scattering channel is active since $\pi_{in}$ and $\pi_{out}$ are not parallel to each other. In Fig. 9(b,c), we plot the outgoing polarization-resolved correction factors $C_{\sigma(\pi)\rightarrow\sigma}(\phi)$ and $C_{\sigma(\pi)\rightarrow\pi}(\phi)$ as well as the total self-absorption correction factors $C_{\sigma(\pi)}(\phi)$ calculated from the above equation.

In the main text, we divided the experimental data by $C_{\sigma(\pi)}(\phi)$  instead of imposing its effect on theory. By this, we remove the effect of self-absorption, and present the azimuthal dependence arising purely from charge/spin/orbital excitations.\\

\section{Appendix D: Raw RIXS spectra of AF-NCO and the fitting procedure}

\begin{figure}
\includegraphics[width=\columnwidth]{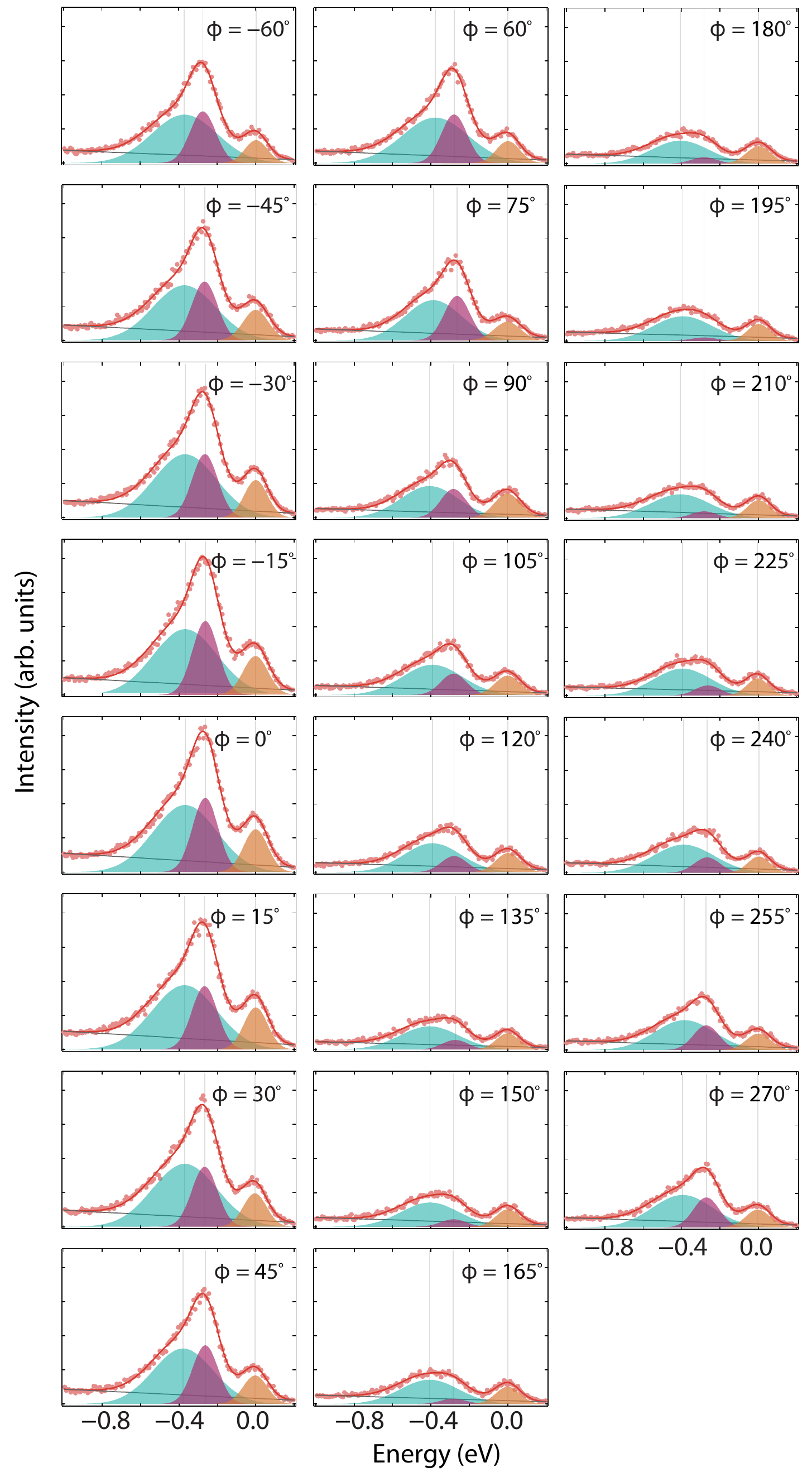}
\caption{\label{fig10} Low-energy RIXS spectra of AF-NCO measured with $\sigma$ incoming polarization under azimuthal rotation.}
\end{figure}

Figure 10 includes the complete azimuthal series of low energy RIXS spectra of AF-NCO before the self-absorption correction. We note that the total acquisition time for a full azimuthal series is comparable to that of outgoing polarization-resolved RIXS experiments, considering the current efficiency of the outgoing polarization filter \cite{Braicovich_thesimultaneous_2014}. The spectra were measured with $\sigma$ incoming polarization. Overall, the intensity is enhanced near $\phi=0^{\circ}$ and suppressed near $\phi=180^{\circ}$, which is the expected behavior in the presence of the self-absorption effects (see Fig. 8).

We fit the spectrum with a linear background and three Gaussian peaks. During the fitting process, we restrict the width of each peak to be constant across all the spectra for different $\phi$. Spectra at all different azimuthal angles can be well fit with $\Delta E=0$, $0.27\pm0.02$, $0.39\pm0.02$ eV, corresponding to elastic scattering, magnon excitations, and bimagnon excitations respectively.

Even without self-absorption correction, we can capture the difference in the azimuthal dependence of each peak. For example, the magnon excitation intensity is almost completely suppressed near $\phi=180^{\circ}$, while a substantial portion of intensity remains in case of elastic scattering and bimagnon scattering. In contrast to the rapid and monotonic suppression of the magnon excitation intensity from $\phi=0^{\circ}$ to $\phi=180^{\circ}$, the intensity of elastic scattering in the raw data is almost constant from $\phi=90^{\circ}$ to $\phi=270^{\circ}$, which can be interpreted as a result of the interplay between the self-absorption correction factor (Fig. 8) and the intrinsic modulation of the cross-section (Fig. 4).

\section{Appendix E: Doping and Azimuthal dependence of the additional peak}

\begin{figure}
\includegraphics[width=\columnwidth]{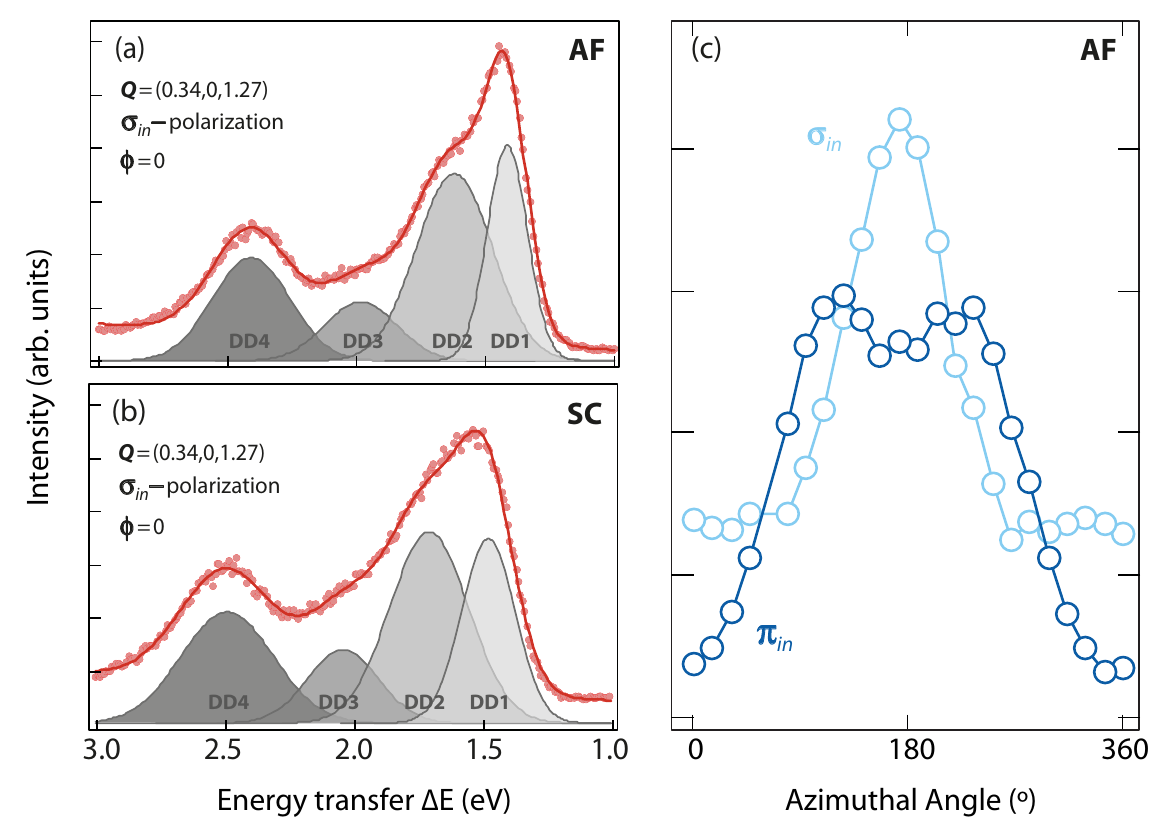}
\caption{\label{fig11} (a),(b) $\textit{dd}$ excitations region of AF-NCO and SC-NCO. Data are measured at $\phi=0^{\circ}$ with incoming $\sigma$ polarization. (c) Azimuthal dependence analysis of 1.95 eV peak in AF-NCO.}
\end{figure}

In Fig. 11, we display the RIXS spectra of high energy $dd$-excitation in AF-NCO and SC-NCO. The additional peak (marked as DD3) near the energy transfer $\sim$2 eV equally appears in both samples with similar intensity. This indicates that an oxygen annealing process marginally affects the peak, except for the slight broadening observed in all $dd$-excitation peaks. Figure 11(c) shows the azimuthal dependence of this peak in AF-NCO. The systematic and non-trivial azimuthal dependence indicates that the peak emerges from a well-defined (yet unknown) excitation, rather than the random defects as previously suggested \cite{moretti_sala_energy_2011}.

\end{document}